\documentclass[preprint,superscriptaddress,nofootinbib]{revtex4}
\usepackage{doi}
\usepackage{hyperref}
\usepackage{bbm}
\usepackage{amsfonts}
\usepackage{mathrsfs}

\usepackage{amsmath,amssymb}

\usepackage{epsfig}
\usepackage{graphicx}               
\usepackage{url}
\usepackage{hyperref}
\usepackage{float}
\usepackage{pstricks}
\usepackage{color}

\newcommand{\nc}{\newcommand}

\nc{\beq}{\begin{equation}}
\nc{\eeq}{\end{equation}}
\nc{\beqa}{\begin{eqnarray}}
\nc{\eeqa}{\end{eqnarray}}
\nc{\bea}{\begin{eqnarray}}
\nc{\eea}{\end{eqnarray}}
\nc{\barray}{\begin{eqnarray}}
\nc{\earray}{\end{eqnarray}}
\nc{\barrayn}{\begin{eqnarray*}}
\nc{\earrayn}{\end{eqnarray*}}
\nc{\ra}{\rightarrow}
\nc{\lsim}{\begin{array}{c}\,\sim\vspace{-21pt}\\< \end{array}}
\nc{\gsim}{\begin{array}{c}\sim\vspace{-21pt}\\> \end{array}}
\nc{\Tr}{{\rm Tr}}
\nc{\slsh}{\slash\hspace*{-0.22cm}}

\def\be{\begin{equation}}
\def\ee{\end{equation}}
\def\bea{\begin{eqnarray}}
\def\eea{\end{eqnarray}}
\def\bit{\begin{itemize}}
\def\eit{\end{itemize}}

\nc{\infinity}{\infty}
\nc{\mc}{\mathcal}
\nc{\M}{\mathcal{M}}

\begin{document}

\title{Indirect Detection Signatures for the Origin of Asymmetric Dark Matter}

\author{Yue Zhao}
\affiliation{Stanford Institute of Theoretical Physics, Physics Department,\\
Stanford University, Stanford, CA 94305, USA}
\author{Kathryn M. Zurek}
\affiliation{Michigan Center for Theoretical Physics, University of Michigan, Ann Arbor, MI 48109, USA}

\begin{abstract}
We study the decay signatures of Asymmetric Dark Matter (ADM) via
higher dimension operators which are responsible for generating the
primordial dark matter (DM) asymmetry.  Since the signatures are
sensitive both to the nature of the higher dimension operator
generating the DM asymmetry and to the sign of the baryon or lepton
number that the DM carries, indirect detection may provide a window
into the nature of the mechanism which generates the DM asymmetry.
We consider in particular dimension-6 fermionic operators of the
form ${\cal O}_{ADM} = X {\cal O}_{B-L}/M^2$, where ${\cal O}_{B-L}
=  u^c d^c d^c,~\ell \ell e^c,~q \ell d^c$ (or operators related
through a Hermitian conjugate) with the scale $M$ around or just
below the GUT scale.  We derive constraints on ADM particles both in
the natural mass range (around a few GeV), as well as in the range
between 100 GeV to 10 TeV. For light ADM, we focus on constraints
from both the low energy gamma ray data and proton/anti-proton
fluxes. For heavy ADM, we consider $\gamma$-rays and
proton/anti-proton fluxes, and we fit $e^+/e^-$ data from AMS-02 and
H.E.S.S. (neglecting the Fermi charged particle fluxes which
disagree with AMS-02 below 100 GeV). We show that, although the best
fit regions from electron/positron measurement are still in tension
with other channels on account of the H.E.S.S. measurement at high
energies, compared to an ordinary symmetric dark matter scenario,
the decay of DM with a primordial asymmetry reduces the tension.
Better measurement of the flux at high energy will be necessary to
draw a definite conclusion about the viability of decaying DM as
source for the signals.
\end{abstract}

\preprint{MCTP-14-02}

\maketitle

\tableofcontents

\section{Introduction}
\label{sec:introduction}

Asymmetric Dark Matter (ADM) is a compelling alternative to WIMP
models of dark matter (DM) with thermal freeze-out.  In these models
the DM density is set by its particle-anti-particle asymmetry,
similar to the baryon asymmetry, rather than by its annihilation
cross-section.  While the idea that DM may carry a particle
asymmetry has existed in the literature for a long
time~\cite{Hut:1979xw,Nussinov:1985xr,Dodelson:1989cq,Gelmini:1986zz,Barr:1990ca,Barr:1991qn,Kaplan:1991ah,Kuzmin:1996he},
it has only been relatively recently that robust classes of models
based on higher dimension operators were
introduced \cite{Kaplan:2009ag}.

The ADM operators communicate an asymmetry between the DM and
visible sectors, and have the advantage that they naturally decouple
at low energies, leading to conserved baryon and DM asymmetries
separately in the two sectors late in the Universe. These operators
take on the form \beq {\cal O}_{ADM} = \frac{{\cal O}_{B-L} {\cal
O}_X}{M^{n+m-4}}, \label{ADMops} \eeq where ${\cal O}_{B-L}$ has
dimension $m$ and ${\cal O}_X$ has dimension $n$. By sharing a
primordial asymmetry between the two sectors, the models naturally
realize the relationship $n_X - n_{\bar X} \sim n_b - n_{\bar b}$.
Since the observed baryon to DM energy density is $\rho_{DM}/\rho_b
\sim 5$, this implies the natural mass scale of ADM is $\sim 5
\mbox{ GeV}$.\footnote{The DM may, however, be heavier if new
$X$-violating interactions are present to deplete the $X$-asymmetry
in comparison to the baryon asymmetry.  We discuss this case further
below.} For a review and list of references of DM models employing
higher dimension operators, see \cite{Zurek:2013wia}.

As outlined in \cite{Zurek:2013wia}, for such higher-dimension ADM
models, there are two basic categories of models.  In the first
class a primordial matter anti-matter asymmetry is shared between
the DM and visible sectors via interactions that are mediated by
heavy particles that become integrated out as the temperature of the
Universe drops
\cite{Cai:2009ia,Shelton:2010ta,Haba:2010bm,McDonald:2011zza,Falkowski:2011xh}.
Such scenarios give rise to DM particles whose relic abundance
carries the same baryon or lepton number as visible particles. The
second category generates opposite charge asymmetries for the SM and
DM sectors via non-equilibrium processes
\cite{Farrar:2005zd,Gu:2007cw,Gu:2009yy,An:2009vq,An:2010kc,Gu:2010ft,Hall:2010jx,Kaplan:2011yj,Bell:2011tn,Cheung:2011if,MarchRussell:2011fi,Graesser:2011vj,Davoudiasl:2010am,Davoudiasl:2011fj,Blinov:2012hq,Huang:2013xfa}.
In this case, the DM particles naturally carry opposite
baryon/lepton numbers relative to SM particles in our Universe.

Examples of operators which may transfer an asymmetry between
sectors are \beq {\cal O}_{B-L} = L H,~~U^c D^c D^c,~~Q L D^c,~~~L L
E^c, \label{RPV} \eeq where $L$ is the chiral supermultiplet of a SM
lepton doublet, $H$ is the Higgs doublet, $U^c,~D^c$ are
right-handed anti-quarks, $E^c$ is a right-handed charged
anti-lepton, and $Q$ is a quark doublet. In the context of
supersymmetry, these operators are $R$-parity violating, and having
the simplest interaction with the DM $X$, the simplest ADM
interactions take the form \beq W_{\rm ADM} = X L H,~~\frac{X U_i^c
D_j^c D_k^c}{M_{ijk}},~~\frac{X Q_i L_j D_k^c}{M_{ijk}},~~~\frac{X
L_i L_j E_k^c}{M_{ijk}}, \label{basicADM} \eeq where now we have
explicitly included a flavor index $i,~j,~k$ on the generic scale of
the operator $M$.

In the context of supersymmetry, the ADM particle is stabilized by
$R$-parity.  On the other hand, the analogue fermionic operators, of
the form\footnote{We do not include other choices of Lorentz
structures for these 4-fermion interactions since they do not make a
substantial difference in the indirect detection signals.} \beq
{\cal O}_{\rm ADM} = X \ell H,~~\frac{X u_i^c d_j^c
d_k^c}{M_{ijk}^2},~~\frac{X q_i \ell_j d_k^c}{M_{ijk}^2},~~~\frac{X
\ell_i \ell_j e_k^c}{M_{ijk}^2}, \label{fermionADM} \eeq may also
share a primordial between the two sectors.  To distinguish from
superpotential multiplets in SUSY, we use lower case letters for the
SM fermionic fields in the Lagrangian, and to label the operator
conveniently, we use the SM part of the operator as a subscript. For
example, we label $\frac{X u_i^c d_j^c d_k^c}{M_{ijk}^2}$ as
$\mathcal {O}_{UDD}$. When working with a non-holomorphic
Lagrangian, instead of a superpotential, many more possibilities
arise, such as \beq {\cal O}_{\rm ADM} = \frac{X d_i^c
u_j^{c\dagger} e_k^{c \dagger}}{M_{ijk}^2},~\frac{X q_i
\ell_j^\dagger u_k^c}{M_{ijk}^2},~ \frac{X q_i d_j^{c \dagger}
q_k}{M_{ijk}^2}. \eeq The effective baryon or lepton number of the
DM (which is defined as being opposite of the $B-L$ charge carried
by ${\cal O}_{B-L}$) in each of the operators differs.  Both types
of operators may be easily UV completed, and the flavor structure
depends on the UV completion. For example, $\frac{X \ell_i \ell_j
e_k^c}{M_{ijk}^2}$ can be obtained by the Lagrangian $\mathscr{L}
\supset y_{i} X \ell_i \Phi + y'_{jk}\Phi^\dag \ell_{j} e^c_k $,
where $i,j,k=1,2,3$ for 3 generations and $\Phi$ is a heavy scalar
field in fundamental representation of $SU(2)_W$. If $y_{i}=y'_{jk}$
for all $i,j$ and $k$, we obtain a universal flavor structure for
$\frac{X \ell_i \ell_j e_k^c}{M_{ijk}^2}$.\footnote{We emphasize
that one can UV complete this operator in another way, i.e.
$\mathscr{L} \supset y_{1,i} X e^c_i \Phi + y_{2,jk}\Phi^\dag
\ell_{j} \ell_{k} $.  In this case, $\Phi$ is a heavy scalar field
but a singlet in $SU(2)_W$.  Since $\ell_{j}$ and $\ell_{k}$ have to
contract by an anti-symmetric tensor in the $SU(2)_W$ basis, they
must be in different generations. A similar subtlety also occurs for
the $\frac{X u_i^c d_j^c d_k^c}{M_{ijk}^2}$ operator.}

While these operators induce an asymmetry in the two sectors, they
also cause the fermionic $X$ to decay. If its abundance has not been
cosmologically depleted in the early Universe, and $M$ is a high
scale, the decay lifetime can be long.  Assuming the heavy mediator
is a scalar field, {\em i.e.} in the form of the effective operators
in Eq.~(\ref{fermionADM}), the decay lifetime is approximately \beq
c \tau \simeq \frac{6144 \pi^3 M^4}{C_{color} C_{flavor} C_{SU(2)_W}
m_X^5}\simeq 3.9\times10^{26}\ \textrm{s}\ (\frac{M}{10^{13}\
\textrm{GeV}})^4(\frac{20\
\textrm{GeV}}{m_X})^5\frac{1}{C_{color}}\frac{1}{C_{flavor}}\frac{1}{C_{SU(2)_W}}.\label{eq::DecayLT}
\eeq Here $C_{color}$, $C_{flavor}$ and $C_{SU(2)_W}$ indicate the
constants introduced from color, flavor and weak isospin
combinations in the final states. 

Observations of the DM decay products in high energy gamma rays and
in charged particles (electrons, positrons and anti-protons) thus
will constrain $M$.  As we will show, if $M \gtrsim 10^{13}$ GeV,
these lifetimes are on the order of current constraints, and their
decay may be detectable both in photons and in charged cosmic ray
byproducts. Similar decay signatures have also been studied in many
other contexts. (Please see \cite{Fan:2010yq} and the references
therein for a review.)  As pointed out in
\cite{Nardi:2008ix,Arvanitaki:2008hq}, current constraints from
indirect detection implies a suppression scale around the GUT scale
if weak scale DM decays through dimension 6 operators.  Most
studies, however, have mainly focused on symmetric DM. In this
paper, we focus on the asymmetric DM scenario, and, as we will see,
the sign of the effective DM baryon or lepton number substantially
affects the results. Refs.~
\cite{Nardi:2008ix,Chang:2011xn,Masina:2011hu,Masina:2013yea,Feng:2013vva,Masina:2012hg}
also studied scenarios where DM particles decay asymmetrically. In
these studies, however, the operators which induce DM decay may not
be those which are responsible for generating the asymmetry in DM
sector as in ordinary ADM models. In Ref.~\cite{Feldstein:2010xe},
the authors briefly mentioned the possibility of ADM decay induced
by the operators in ordinary ADM models, though they were mainly
focused on the neutrino fluxes induced from other operators. In
addition, the studies mentioned above only focused on a few specific
decay channels, while we carry out a comprehensive study of ADM
decay through various operators.

The goal of this paper is two-fold.  First, we aim to study the
constraints from photons in the galactic center and diffuse extra
galactic background on the scale $M$ in Eq.~(\ref{fermionADM}) from
fermionic ADM, assuming the fermionic ADM composes all (or most of)
the DM.  We do this both for ADM in its natural mass window (from a
few GeV up to approximately 20 GeV), and for ADM with a heavier mass
near the weak scale. Second, we study models of ADM that may
generate part or all of the charged cosmic ray signals observed by
PAMELA, AMS-02 and H.E.S.S., consistent with the flux of
anti-protons in the Universe.

There are many ADM models where the DM mass is much heavier than a
few GeV.  In this case a mechanism must be present to reduce the DM
number density relative to the baryon number density.  This can be
achieved, for example, by inducing DM/anti-DM oscillations that wash
out the asymmetry so that subsequent annihilations can reduce the DM
number density.  In this case the DM is not asymmetric from an
indirect detection point of view.  It is not difficult, however, to
build a model where the DM is electroweak scale while retaining its
asymmetry throughout the history of the Universe.  One
straightforward way to achieve this is to assume a non-zero
primordial baryon/lepton (B/L) number in a parent particle (such as
the state integrated out to generate the operators
Eq.~(\ref{ADMops})) which subsequently decays with different
branching fractions to the DM and the visible sectors.  Such a
scenario is discussed in
\cite{Thomas:1995ze,Kitano:2004sv,Unwin:2012rp}.  As long as the DM
and SM sectors are never in thermal equilibrium after decay of the
heavy particles, the DM mass can be tuned to any value by changing
the primordial asymmetry. In addition, the asymmetry can be diluted
later in the Universe through a DM-number violating process (such as
annihilation) which washes out the asymmetry; we present such a
model in Appendix \ref{sec:HeavyADM}.

The outline of this paper is as follows.  We first discuss the
details of the operators we study and specify the flavor structure
for each operator in Sec.~\ref{sec:operators}. Then, in Sec.
\ref{sec:Gamma}, we provide details of the gamma ray flux
calculation, for both the galactic and diffuse extra-galactic gamma
rays.  In Sec. \ref{sec:chargedCRDM} we focus in detail on
charged cosmic ray fluxes, both electron/positron and
proton/anti-proton fluxes. In \ref{sec:ADMresult}, we discuss the
results for light and heavy ADM scenarios. For heavy ADM, we find
the best-fit region for electron/positron fluxes. Gamma ray spectra
and proton/anti-proton fluxes are used to constrain the parameter
space for both light and heavy ADM scenarios.  Finally we conclude, reviewing our results.

\section{Operators for Asymmetric Dark Matter Decay}
\label{sec:operators}

There are many signatures that can arise from DM decay through the
operators in Eq.~(\ref{fermionADM}).  It is the purpose of this
section to motivate the particular choices of flavor structures in
these operators that we study below.  We do not consider the $X \ell
H$ operator, which is marginal and will lead to rapid DM decay.


As discussed in the introduction, in most ADM models, the mass of
the DM particle is naturally $1\sim 20$ GeV.  The DM may,
however, be heavier.  Besides the possibility of a primordial
asymmetry in the heavy particles which induce the asymmetry in DM/SM
sectors through decay
\cite{Thomas:1995ze,Kitano:2004sv,Unwin:2012rp}, we provide an
alternative option in Appendix \ref{sec:HeavyADM}. There we build a
toy model of thermal ADM where the DM is heavier, which occurs if
some $X$-violating interaction (mediating annihilations) is in
thermal equilibrium when the temperature $T \sim m_X$.  In this
case, the DM number density is suppressed by a Boltzmann factor
$e^{-m_X/T_{fo}}$, where $T_{fo}$ is the temperature at which
freeze-out of the $X$-violating interactions occurs.  Since we focus
on the phenomenology of ADM decay, we treat the DM mass as a free
parameter, and we divide our discussion into two parts.  We will
first focus on the natural mass range of ADM models, i.e. $3\
\textrm{GeV} < m_{DM} < 20\ \textrm{GeV}$. Then we study the case
that $100 \mbox{ GeV} \lesssim m_X \lesssim 10 \mbox{ TeV}$.  We
emphasize that this latter case, while motivated by models of ADM,
may arise in many GUT-inspired models, such as those explored in
\cite{Nardi:2008ix,Arvanitaki:2008hq,Heckman:2011sw}.

In ADM models, the DM effectively carries non-zero baryon or lepton
number, which may be positive or negative in sign.  The gamma ray
spectra are indifferent to the sign of the baryon or lepton number
of the DM, but it is crucial for the charged cosmic ray
measurements. We will consider both cases in our study.

The flavor structure of each model, on the other hand, is important
for gamma ray observations. The possible flavor structures are many
fold, and, because of the high scale of the operator, unrestricted
by flavor constraints.  For leptons in the final state, the
electron/positron gives a hard spectrum since photons are from FSR,
while the photon spectrum from tau decay is softer because of the multi-step nature of tau decay. Further, the injection spectra of the
electron/positron can directly affect both electron/positron fluxes
on the Earth and the gamma ray flux from Inverse Compton (IC)
processes. Thus the flavor structure in the lepton sector has large
effects on observations. For operators with colored particles in the
final states, the third generation is special in a two-fold manner.
First, its large mass can affect the kinematic distributions of
final state particles. In the low mass region, i.e. $3\ \textrm{GeV}
< m_{DM} < 20\ \textrm{GeV}$, the $b$-quark mass is important for
kinematics, while in the high mass region, {\em i.e.} $100 \mbox{
GeV} \lesssim m_X \lesssim 10 \mbox{ TeV}$, the top quark mass is
important when the DM mass is a few hundred GeV.  Second, the third
generation quarks have different hadronization and decay products
compared to the first two generations.  For example, the top quark
decay can contribute hard leptons in the final state.

For a light ADM mass, we treat most of the operators as flavor
universal. As an example to explicitly show how flavor affects the
observations, we take the $\mathcal {O}_{UDD}$ operator and specify
its decay products in two scenarios, {\em i.e.} light quarks only
($\mathcal {O}_{UDD_L}$) and the heaviest quarks kinematically
accessible only ($\mathcal {O}_{UDD_H}$).  As we discussed
previously, however, flavor is more important for the heavy ADM
scenario since electron/positron fluxes are involved.  Thus for
heavy ADM we study all operators in the two extremal limits, {\em
i.e.} the lightest generation or the heaviest flavor kinematically
accessible. One consequence of this flavor choice is that the decay
through $\mathcal {O}_{LLE}$ is flavor symmetric, since, due to
charge conservation, there must be two oppositely charged leptons in
the final states. However, if the charged leptons in the final
states are not in the same generation, the asymmetric nature of the
decay may become phenomenologically apparent.  As an example, to
highlight this unique feature of ADM models, we study one more decay
channel for $\mathcal {O}_{LLE}$, $DM\rightarrow
e^{\pm}+\tau^\mp+\nu (\bar{\nu})$.

In addition to the flavor structure of operators, each class of
operators has several variations.  As mentioned above, the Lorentz
structure of the four Fermi interaction is not important for the
indirect detection signals, so that we focus on the contraction
integrating out the scalar particle which generates the four Fermi
interaction in Eq.~(\ref{fermionADM}).  Further, one can change the
operators by taking charge conjugation on part of the operator. For
example, with a small change of the field content to preserve gauge
symmetry, $O_{QLD}$ we may have not only $\frac{X q_i l_j
d_k^c}{M_{ijk}^2}$, but also $\frac{ X l_i q_j^\dagger u_k^{c
\dagger}}{M_{ijk}^2}$.  However, such changes leave the indirect
detection signals essentially unchanged, so that we do not study
this variation of the operators further.\footnote{Since the d-quark
is replaced by a u-quark, the FSR spectrum may change by a small
amount due to the different charges of u and d quarks. However, this
change is negligible since the dominant photons are from
hadronization.}

Finally, one can also change the $SU(2)_W$ field content of the
operator. For example, $O_{QLD}$ can be changed to $\frac{ X d_i^{c}
e^{c\dag}_j u^{c\dag}_k}{M_{ijk}^2}$.  The new operator eliminates
the hard neutrino, and only a charged lepton appears in the final
state. This change impacts both the gamma ray flux and the
electron/positron flux.  We will take
this operator as an example to illustrate the differences induced by
this modification.

We summarize the combinations of operators we consider in
Table~\ref{structures}.

\begin{table}
\begin{tabular}{|c|c|c|} \hline
Operator & light ADM & heavy ADM \\ \hline \hline
$\ell \ell e^c$ & flavor universal & $e^+ + e^- +\nu$ \ or\ $\nu + \nu +\bar{\nu}$\\
&&$\tau^+ + \tau^- +\nu$ \ or\ $\nu + \nu +\bar{\nu}$\\
&&$e^+ + \tau^- +\nu$ \ or\ $\nu + \nu +\bar{\nu}$\\ \hline
$q \ell d^c$ & flavor universal & $e^- + u+\bar{d}$ \ or \ $\nu + d +\bar{d}$\\
&& $\tau^- + t + \bar{d}$ \ or \ $\nu + b +\bar{b}$\\ \hline
$d^{c} u^{c\dag} e^{c\dag} $ & similar to $q \ell d^c$ & $e^- + u+\bar{d}$ \\
& not discussed &$\tau^- + t + \bar{d}$ \\ \hline
$u^c d^c d^c$  & $u + d + s$ & $u + d + s$\\
 &$c + b + s$& $t + b + s$\\ \hline
\end{tabular}
\caption{ADM decay operators and the flavor structures of their
decays, for $\ell \ell e^c$, $q \ell d^c$, $d^{c} u^{c\dag}
e^{c\dag} $ and $u^c d^c d^c$.  For light ADM decay, we choose
flavor universal decay for $\mathcal {O}_{LLE}$ and $\mathcal
{O}_{QLD}$, while for $\mathcal {O}_{UDD}$ we choose two extremal
limits as an illustration. For the gamma ray flux, $\mathcal
{O}_{UDE}$ and $\mathcal {O}_{QLD}$ are very similar, so that we we
will not study $\mathcal {O}_{UDE}$ in the low mass scenario.  For
heavy ADM decay, the flavor structure is important for the charged
cosmic ray study.  We divide our study into two extremal limits
(decay to lightest generation only, and decay to heavy generation),
with an additional flavor asymmetric choice for $\mathcal
{O}_{LLE}$, which highlights the capabilities of indirect detection
to tag ADM signatures. In the table, we do not distinguish the
flavor of neutrinos, and we present only the decay products for ADM
carrying positive B or L number, though we consider ADM with both
positive and negative B(L) number in our study. } \label{structures}
\end{table}

\section{Photons from Dark Matter Decay}
\label{sec:Gamma}

Photons can be produced in many ways in DM decay processes. Charged
particles in the final state can produce photons through
bremsstrahlung.  If there are colored particles in the final states,
hadronization produces $\pi^0$s, which will
decay to photons.  Since these photons are produced directly
from the primary decay process, they are generically energetic. We
will call photons from either bremsstrahlung or hadronic decays $\textrm{FSR}\gamma$. The other important source
of photons is Inverse-Compton (IC) scattering between
energetic electrons/positrons and galactic ambient light, which is
mainly CMB photons and starlight. Since the galactic ambient light
has very low energy, these IC photons are generically much softer
than FSR photons.

In this section, for completeness, we overview the gamma ray spectra
from these sources.  We first focus on the gamma ray spectrum from
the DM halo in our galaxy, then we will discuss the diffuse gamma
ray background.  We summarize the data and statistical procedure we
used in our analysis.

\subsection{Photon Flux from DM Decay} \label{sec:DMdecay}

\subsubsection{Galactic DM Halo} \label{sec:Halo}

The galactic DM halo provides a promising place to look for the
gamma ray flux produced through DM decay processes, where the FERMI
collaboration has released the sky map of the gamma ray measurement
up to a few hundred GeV \cite{FermiLAT:2012aa,Ackermann:2012qk}.
Electrons/positrons propagating in the galaxy scatter with
starlight, as well as infrared and CMB photons to produce
Inverse-Compton photon.  The spectrum from IC scattering, especially
in the inner galaxy, depends strongly on details of the galaxy, such as
starlight spectrum and distribution. To avoid introducing large
uncertainties, we do not consider the IC spectrum and only focus on
the FSR$\gamma$ for the galactic halo constraints.

The flux of photons from DM decay in our galaxy can be
written as
\begin{equation}
\frac{d J_\gamma}{d E d\Omega}= \frac{1}{4\pi \tau_{DM}
m_{DM}}\frac{d N_{\gamma}}{d E} \int_{l.o.s.} ds\ \rho_{DM}(r)
\end{equation}
where the integral is along the line of sight, $\frac{d
N_\gamma}{d E}$ is the gamma ray spectrum from ADM decay, and $\rho_{DM}(r)$ is the DM profile in our galaxy.  We choose an
$NFW$ profile,
\begin{equation}
\rho_{DM}(r) = \rho_s (\frac{r_s}{r}) (1+\frac{r}{r_s})^{-2}
\label{Eq:NFW}
\end{equation}
with $r_s = 24.42\ \textrm{kpc}$ and $\rho_s = 0.184\
\textrm{GeV}/\textrm{cm}^3$.  To get the gamma ray spectrum from
DM decay, i.e. $\frac{d N_\gamma}{ d E}$, we use MadGraph
to generate parton level events, and use PYTHIA to shower and
hadronize the events.

\subsubsection{Extra-galactic $\gamma$-ray} \label{sec:exG}

In addition to the galactic halo, the gamma ray flux from the decay
or annihilation of DM particles in the early Universe can propagate
to the Earth and contribute as a diffuse extra-galactic gamma ray
background. The measurement of the diffuse extra-galactic gamma ray
spectrum is provided by FERMI in \cite{Abdo:2010nz}, and provides a particularly important constraint on DM decay. The ratio of
extra-galactic gamma ray flux from DM decay, $\Phi_{exG-\gamma}$, to
the galactic halo gamma ray flux, $\Phi_{halo}$, can be estimated as,
\beq\frac{\Phi_{exG-\gamma}}{\Phi_{halo}}\sim
\frac{\rho_{cosmo}R_{cosmo}}{\rho_\odot R_\odot}\sim 1, \eeq where
$\rho_{cosmo}$ is the average DM energy density in the Universe,
$R_{cosmo}$ is the size of the Universe, $\rho_\odot$ is the local
DM energy density and $R_\odot$ is the distance from the solar
system to the galactic center. Due to this numerical coincidence,
the constraints from the diffuse extra-galactic gamma ray flux are
comparable to the constraints from the galactic halo.

There are again two dominant contributions to extra-galactic gamma
rays, one from FSR$\gamma$ and the other from
scattering between hard electron/positrons produced from the decay
and the soft photon background.  Unlike in the galaxy, the IC
scattering is dominated by scattering off CMB photons. Since
the uncertainty is rather small in this case, we will include the IC
contribution to the diffuse extra-galactic gamma ray flux.

For the photons produced directly from DM decay, the spectrum can be
calculated by properly redshifting the photon injection spectrum at
any redshift $z$. High energy photons can be absorbed in
a cosmological length. The dominant absorption is caused by the
scattering with CMB photons. This is only important, however, for
extremely high energy photons. For the energy range we consider in
this paper, {\em i.e.} $E_\gamma$ smaller than few $TeV$, the
absorption is negligible. Given an injection spectrum from DM decays at redshift $a = 1/(1+z)$, {\em i.e.}
$\frac{dN_{\gamma,\textrm{FSR}}}{dE_\gamma(a)}$, the flux of photons
is
\begin{eqnarray}
\frac{d^2 \Phi_{\gamma,EG,\textrm{FSR}}}{d\Omega dE_\gamma} =
\frac{c\ \Omega_{DM}\rho_{c}}{4\pi \tau M_{DM}}\int^1_0
\frac{da}{a^2}
\frac{1}{H_0\sqrt{\Omega_\Lambda+\Omega_m/a^3}}\frac{dN_{\gamma,\textrm{FSR}}}{dE_\gamma(a)}.
\label{Eq:exGFSR}
\end{eqnarray}
We take $\Omega_m+\Omega_\Lambda\simeq 1$ and
$\Omega_{DM}\rho_{c}\simeq 1.3 \times
10^{-6}\textrm{GeV}/\textrm{cm}^3$, when calculating the gamma ray flux
from prompt photons.

To estimate the gamma ray flux from the IC scattering between high
energy electrons/positrons and CMB photons, we closely follow the
procedure of \cite{Cirelli:2010xx}. For low energy photons in the
CMB, the radiation power and the energy loss coefficient function
are computed in the Thomson limit. This simplifies the calculation.
Further, the mean free path of the electron/positron in the
intergalactic medium is much shorter than the cosmological length,
so that one can approximately treat the IC spectrum as injected
instantaneously, $\frac{dN_{\gamma,IC}}{dE_\gamma(a)}$.  Similar to
Eq. (\ref{Eq:exGFSR}), by properly redshifting the IC spectrum, one
obtains the IC contribution to the extra-galactic gamma ray.

When the DM mass is
small, the IC contribution to the extra-galactic gamma spectrum is
negligible.  However, when the DM is very heavy, {\em e.g.} {\cal O}(TeV), the IC
contribution is dominant.  We will see this explicitly when we
discuss the heavy ADM scenario.

\subsection{Data and Statistical Methodology}\label{sec:data}

\begin{figure}
\begin{center}
\hspace*{0.15cm}
\includegraphics[width=0.95\textwidth]{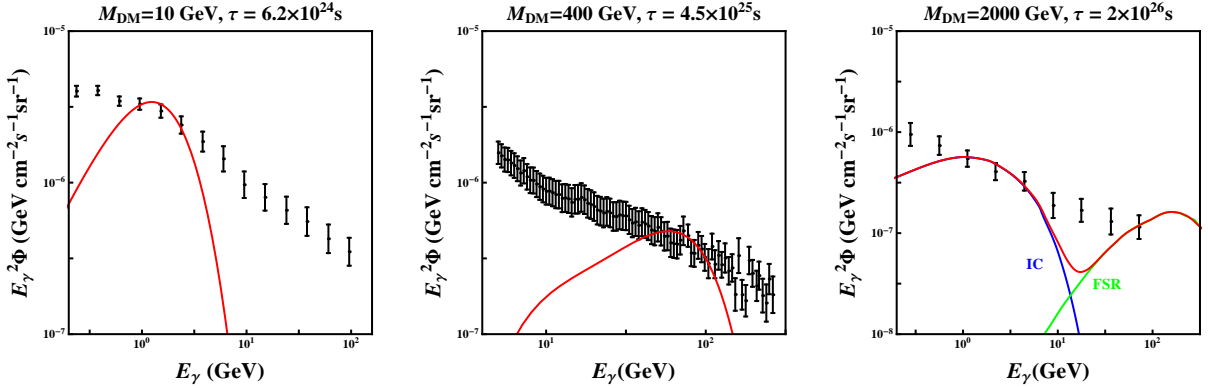}
\caption{From left to right, we show the Fermi galactic low energy
data \cite{FermiLAT:2012aa}, the Fermi galaxy high energy data
\cite{Ackermann:2012qk}, and the Fermi diffuse extragalactic gamma ray data
\cite{Abdo:2010nz}.  We also show the ADM decay spectra through the
$\mathcal {O}_{LLE}$ operator assuming a flavor universal structure,
shown as red curves. For the extragalactic gamma ray flux, when the DM mass
is large, both FSR and IC contributions are important. The decay
lifetime is chosen so that gamma ray from DM decay does not exceed
any bin by $1-\sigma$.} \label{fig:data}
\end{center}
\end{figure}

For the galactic gamma ray spectrum, the FERMI collaboration
provides two sets of measurements which we use.  One is focused on
the low energy regime, ranging from 0.2 GeV to 100 GeV
\cite{FermiLAT:2012aa}. In this measurement, the gamma ray spectrum
is provided on different patches on the sky. We choose the patch of
the full sky without the galactic plane, i.e. $0^\circ\le l \le
180^\circ$ and $8^\circ\le b \le 90^\circ$.  When the DM mass is
small, the low energy measurement is the most sensitive probe.  The
other measurement from the FERMI collaboration is in the high energy
regime, from 4.8 GeV to 264 GeV \cite{Ackermann:2012qk}.  The region
of coverage is the full sky minus the galactic plane while keeping
galaxy center, {\em i.e.} $(|b|>10^\circ)|(l\le 10^\circ)|(l\ge
350^\circ)$. This will be more useful for constraining the heavy ADM
decay scenario. For the diffuse extra-galactic gamma ray spectrum,
we take the most recent published measurement from FERMI
\cite{Abdo:2010nz}. In Fig.~\ref{fig:data}, we overlay all the data
sets we use for our gamma ray analysis.

In this paper we provide the most conservative constraints on the ADM
decay scenario from gamma ray spectra.  We require the flux from ADM
decay does not exceed the central value plus twice the error bar
in any bin, without any assumption about the background flux. One could improve the constraints by subtracting the
astrophysical background, gaining perhaps a factor of a few on the
constraints. This, however, induces larger systematic uncertainties
from the background.  For this reason, we focus on the most
conservative analysis.

\section{Charged Particles From Dark Matter Decay}
\label{sec:chargedCRDM}

In this section, we focus on charged
particle fluxes induced by DM decay.  As noted in the introduction,
unlike in previous studies, the ADM operators we employ, {\em i.e.}
Eq.~(\ref{ADMops}), both generate the DM asymmetry and induce DM
decay, so that in this case the cosmic ray signals are a signature
for the ADM mechanism itself. At minimum, the asymmetry of the DM
impacts signatures through the sign of the baryon or lepton number
that the DM carries, which in turn determines the nature of the
decay products. Since the signatures depend on the B/L sign, we will
consider both cases.  In addition, as usual, the flavor structure of
the operators affects the signatures substantially,
especially for the study of best-fit region for
electron/positron fluxes, as summarized in Table~\ref{structures}.
For the light DM scenario, we study the flavor universal
scenario except for $\mathcal {O}_{UDD}$. For the heavy DM
scenario, we will take two extremal cases in this section -- DM
decaying to the first generation only, or to the third generation
only; other flavor combinations fall between these two choices.  In
addition, for  the $\mathcal {O}_{LLE}$ operator we make another
flavor choice, decay to $e^+ \tau^- \nu$, that highlights the
asymmetric nature of the decay. When DM is a symmetric relic,
generically, one expects the same spectra of electrons and positrons
in the final state.\footnote{There are some special cases where even
symmetric dark matter decay can induce asymmetric electron/positron
spectra. One example is assuming DM is a Majorana fermion with
several different decay channels.  If there is a non-trivial
CP-violating phase, then the electron/positron spectra in the final
states can be different from each other. This scenario is realized
in \cite{Falkowski:2011xh}, though not aimed at inducing DM decay.}
However, this is not necessarily the case for ADM -- since there may
be no hard electrons in the final state, the positron ratio from DM
decay alone can be as high as 1, and since there are no hard
electrons in the final state, the number of hard photons from FSR as
well as IC is reduced. These special features of the ADM scenario
help to reduce the tensions between the AMS-02 anomaly and other
measurements \cite{Masina:2012hg}.

We have already discussed in Sec. \ref{sec:Gamma} the methods that
we use for constraining ADM decay with photons.  Thus in this
section, we will focus on the electron/positron flux and
proton/anti-proton flux, where we provide details on the data we use
and the statistics we apply. In Sec. \ref{sec:ADMresult}, we
present our results by combining all channels for indirect
detection, both gamma and charged cosmic rays.

\subsubsection{e+/e- Fit from AMS-02 and H.E.S.S.} \label{sec:HADMepem}

In 2008, PAMELA \cite{Adriani:2008zr} published their measurements
of the electron/positron fluxes, showing that the positron fraction
rises at energies above few GeV.  Recently AMS-02
\cite{Aguilar:2013qda} confirmed PAMELA's result but with smaller
uncertainties and extending to higher energies. Since ADM decays to
quarks and leptons through the operators in Eq.~(\ref{ADMops}), it
is interesting to see how well the electron/positron flux can be
fitted by these operators. We use AMS-02 data only for our fit in
low energy regime; since AMS-02 is in good agreement with PAMELA, we
do not expect inclusion of the PAMELA data to substantially change
our result. This reduces the uncertainties on combining different
data sets from different experiments.  For the total $e^\pm$ flux
measurement, we fit the AMS-02 and H.E.S.S. data (the latter being
relevant only at the highest energies). We do not include Fermi. The
measurements of Fermi and AMS-02 disagree below 100 GeV so that
including both Fermi and AMS-02 data would give rise to a poor fit.
We have checked that including Fermi instead of AMS-02 data in our
fits does not substantially change our result, since in that case
the fit simply prefers a different astrophysical background. Further
work and measurement will be required to resolve the systematic
difference between Fermi and AMS-02 below 100 GeV.


To obtain the electron/positron fluxes received near the Earth, we
use GALPROP to calculate the propagation \cite{Moskalenko:1999sb}.
We run the 2D mode of the code, which calculates the propagation
equations on $(r,z)$ grid.  We use the same DM distribution profile
applied in previous studies, {\em i.e.} Eq.~(\ref{Eq:NFW}), and we
choose the propagation parameters in a conventional way.  The
diffusion constant $K(E)$ is taken to be $5.8\times 10^{28}(E/4\
{\rm GeV})^{0.33}\ {\rm cm}^2/{\rm s}$, and the root-mean-square of
the magnetic field is modeled by an exponential disk,
\begin{eqnarray}
B_{\rm rms}=B_0\ exp(-(r-R_{\odot})/r_B-|z|/z_B)
\end{eqnarray}
where $B_0=5~\mu$G, $r_B=10 $ kpc and $z_B=2$ kpc.

To estimate how well electron/positron fluxes constrain the decay
lifetime, we carry out a $\chi^2$ fit including an astrophysical
background, which we take to be
\cite{Moskalenko:1997gh,Baltz:1998xv}
\begin{eqnarray}
&\Phi_{e^-}^{(prim)}(E)& = \frac{0.16 e^{-1.1}}{1+11 e^{0.9}+3.2 e^{2.15}}\ ({\rm GeV}^{-1} {\rm cm}^{-2} {\rm s}^{-1} {\rm sr}^{-1})  \nonumber  \\
&\Phi_{e^-}^{(sec)}(E)& = \frac{0.7 e^{0.7}}{1+110 e^{1.5}+ 600 e^{2.9}+580 e^{4.2}}\ ({\rm GeV}^{-1} {\rm cm}^{-2} {\rm s}^{-1} {\rm sr}^{-1})  \\
&\Phi_{e^+}^{(sec)}(E)& = \frac{4.5 e^{0.7}}{1+650 e^{2.3}+ 1500
e^{4.2}}\ ({\rm GeV}^{-1} {\rm cm}^{-2} {\rm s}^{-1} {\rm sr}^{-1})  \nonumber
\end{eqnarray}
where $e=\frac{E}{1 \ {\rm GeV}}$.  To treat the background uncertainties,
we allow variation in both overall normalization and index
of the power law.  More precisely, we take
\begin{eqnarray}
&\Phi_{e^-}(E)& = A_- e^{P_-}(\Phi_{e^-}^{(prim)}(E)+\Phi_{e^-}^{(sec)}(E))   \nonumber \\
&\Phi_{e^+}(E)& = A_+ e^{P_+}\Phi_{e^+}^{(sec)}(E)
\end{eqnarray}
where $0< A_\pm < +\infty$ and $-0.05< P_\pm < 0.05$. To fit the
AMS-02/H.E.S.S. data, we took both the positron ratio and $e^\pm$
total flux with 6 parameters, $A_\pm$, $P_\pm$, $m_{DM}$ and $\tau$.
We only take the bins with energy larger than 10 GeV in order to
reduce the uncertainties from solar modulation. Further, to fit the
total electron/positron flux from H.E.S.S. measurement, we include
the 15\% systematic uncertainty in the energy calibration as
following:

\begin{eqnarray}
\chi_{H.E.S.S.}^2 = min\{\ \sum_i
\frac{(\Phi_i^{DM}(E_i(1+e))-\Phi_i^{exp})^2}{\delta\Phi^2}+\frac{e^2}{\delta
e^2}\  |\ e\ \}
\end{eqnarray}
where the sum runs over all bins in H.E.S.S. data, and we take
$\delta e$ as 15\%. Later, we present the 3-sigma best fit region in
the $(m_{DM}-\tau)$ plane.

To illustrate how well one can fit AMS-02 and H.E.S.S. data, we
choose several benchmark points and show the comparison between the
fit and the data.  For positron ratios, we extend curves beyond
current energy range to show how various models behave as more
AMS-02 data is accumulated. Complete results for different ADM
operators will be shown below, in Sec.~\ref{sec:ADMresult}.

\begin{figure}
\begin{center}
\hspace*{0.35cm}
\includegraphics[width=0.405\textwidth]{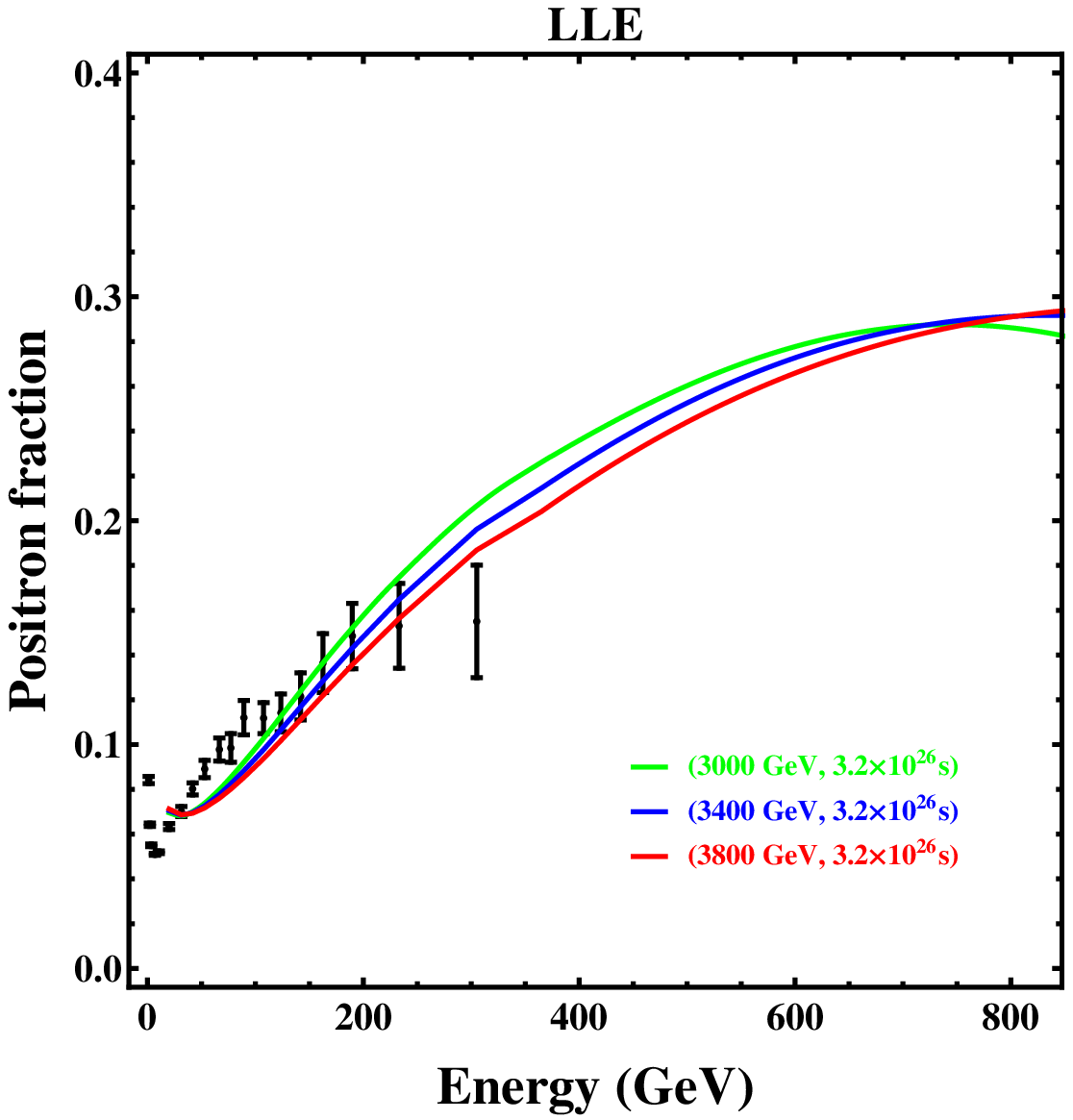}
\hspace*{0.5cm}
\includegraphics[width=0.405\textwidth]{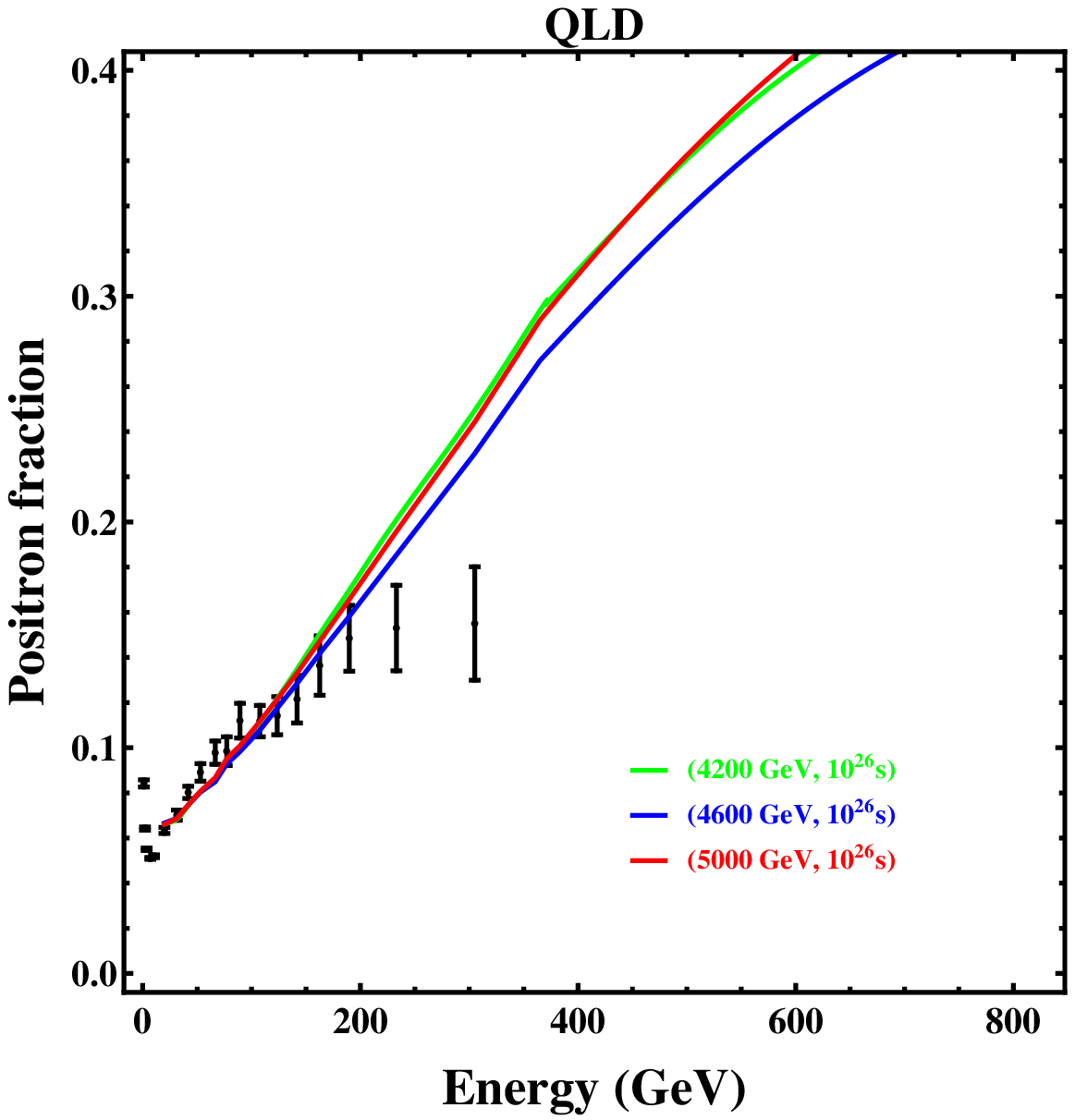}
\includegraphics[width=0.44\textwidth]{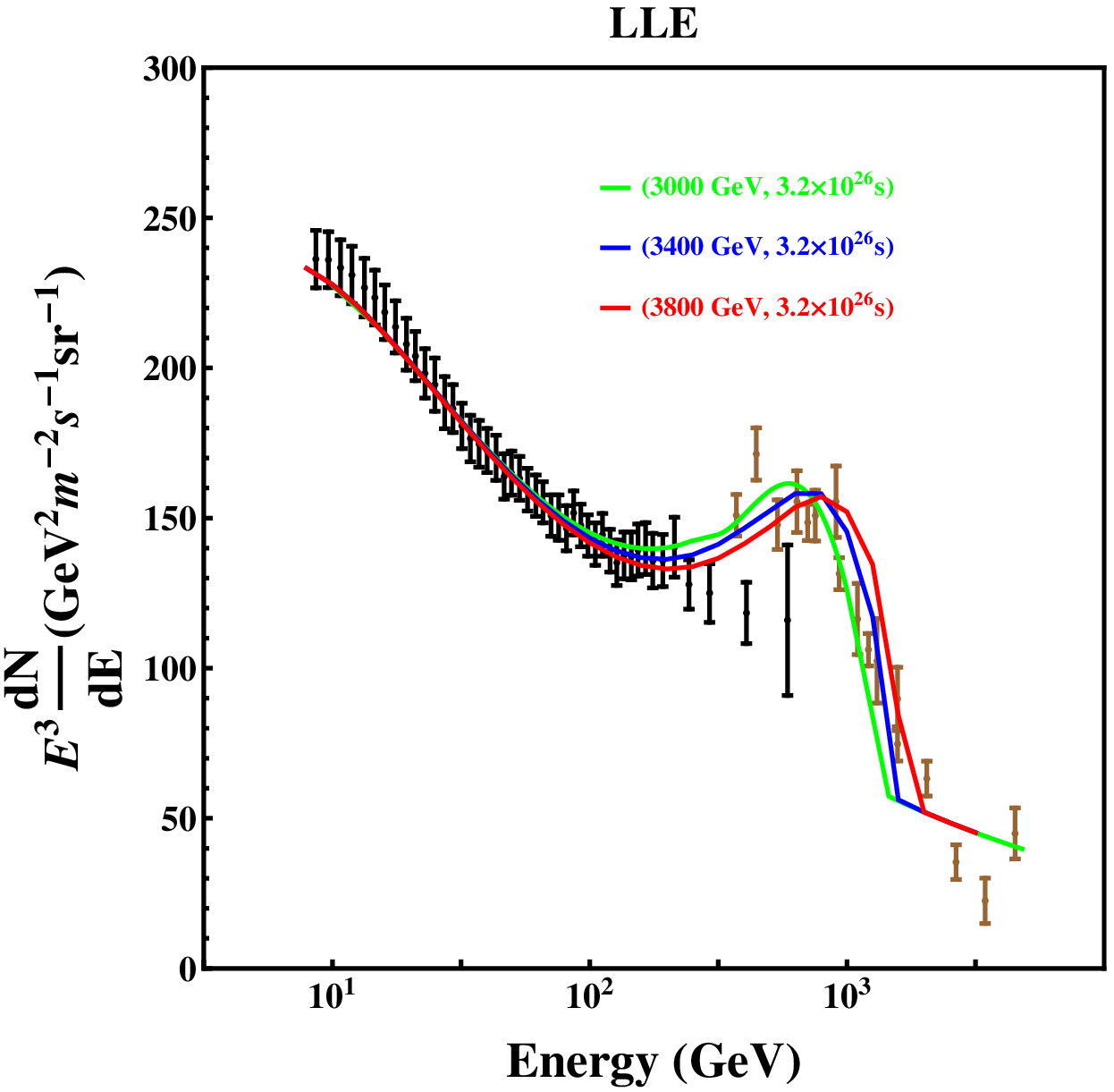}
\includegraphics[width=0.44\textwidth]{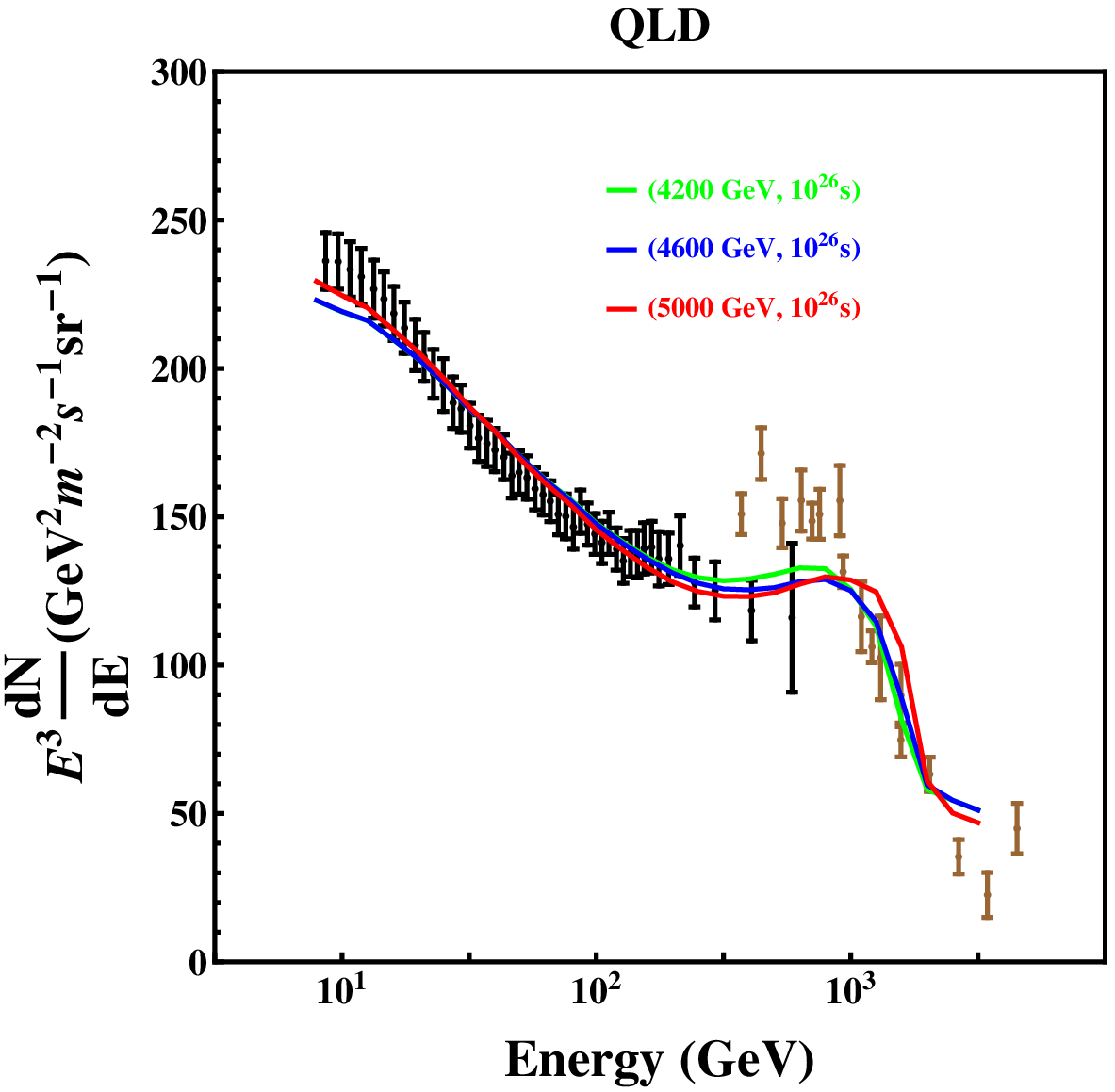}
\caption{Benchmark points of the electron/positron spectra. {\em
Left}: DM decay through the $\mathcal {O}_{LLE}$ operator, with
first generation fermions in the final states only. DM masses
are taken to be 3 TeV, 3.4 TeV and 3.8 TeV, with the decay lifetime fixed at $3.2\times 10^{26}$s. {\em Right:} DM decay through $\mathcal
{O}_{QLD}$, with first generation fermions in the final states only.
DM masses are taken to be 4.2 TeV, 4.6 TeV and 5 TeV, with decay
lifetimes fixed at $10^{26}$s.  Data points are taken from the recent
AMS-02 results \cite{Aguilar:2013qda} and H.E.S.S. measurements
\cite{Aharonian:2008aa,Aharonian:2009ah}. For positron ratios, we
extend curves beyond the current energy range, to show how AMS-02
data might appear at higher energies.} \label{fig:compareepem}
\end{center}
\end{figure}

\subsubsection{Constraints from p+/p- Fluxes} \label{sec:HADMpppm}

For operators we are considering, DM decay products may
include quarks so that modifications of the proton/anti-proton
fluxes are possible.  The best data for the proton flux is from
AMS-02 \cite{Aguilar:proton}, while PAMELA provides the most updated
results for the anti-proton flux and anti-proton/proton ratio
\cite{Adriani:2010rc,Adriani:2011cu}.   For proton/anti-proton
fluxes, the data agrees well with the astrophysical expectation, so
that we use this data to constrain the decay lifetime for each
operator.  Unlike the electron/positron fluxes, the anti-proton flux
is much smaller than the proton flux, with the ratio being $\sim
10^{-4}$ in the energy range of interest. Proton and anti-proton
fluxes are dominantly from the hadronization of colored particles in
the DM decay final states, with the flux of protons comparable to
anti-protons. Thus after adding in the contribution of DM
decay, the anti-proton flux can be changed significantly while the
proton flux remains almost unchanged, implying that the constraint
from anti-proton ratio should be much stronger than that from
proton/anti-proton total flux.

To compute the anti-proton flux as a constraint on ADM decay, we
applied GALPROP to calculate the propagation of the
proton/anti-proton flux, where the parameters are the same as in
Sec. \ref{sec:HADMepem}.  The solar modulation effect is
important in low energy bins.  For the heavy ADM scenario, to reduce the
uncertainties in the solar modulation calculation of the fluxes, we
focus on proton/anti-proton fluxes whose kinetic energy is larger
than 1 GeV. On the other hand, the data in the low energy region is
important for the light ADM scenario. To properly estimate the
constraint on the decay lifetime, we use a force-field approximation to
model the solar modulation:
\begin{eqnarray}
J(E)=\frac{E^2-m^2}{(E+\phi)^2-m^2}J_{IS}(E+\phi)
\end{eqnarray}
where $E$ is the total energy of the proton, $m$ is proton mass.
$J_{IS}$ is the interstellar cosmic ray flux before accounting for
the effect of solar modulation, and $J(E)$ is the cosmic ray flux
after correcting solar modulation effects. $\phi$ is the modulation
parameter which is taken to be 500 MeV.

\begin{figure}
\begin{center}
\hspace*{-0.75cm}
\includegraphics[width=0.40\textwidth]{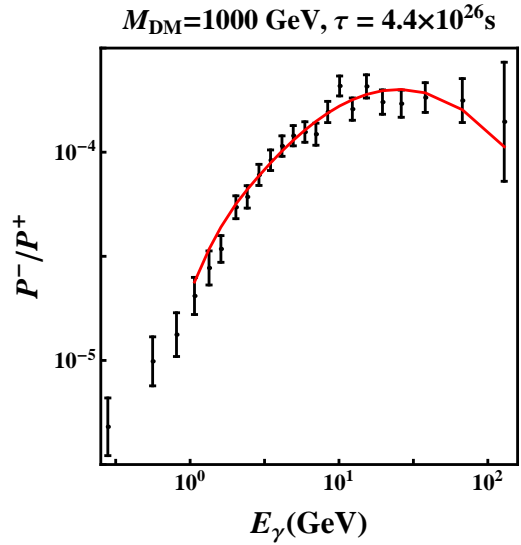}
\caption{Anti-proton to proton flux ratio for a benchmark ADM decay,
adding the DM proton/anti-proton fluxes to the astrophysical
background, and comparing with PAMELA data \cite{Adriani:2010rc}.
The DM mass is 1 TeV, the decay operator $\mathcal {O}_{QLD}$ with
only first generation particles in the final states, and a lifetime
$4.4\times 10^{26}$s.} \label{fig:comparepppm}
\end{center}
\end{figure}

To model the astrophysical background of proton and anti-proton
fluxes, we fit the proton/anti-proton fluxes as  sum of polynomials.
Similar to the electron/positron cases, we allow small variations in
both the overall normalization and the index of the power law, $0<
A_\pm < +\infty$ and $-0.05< P_\pm < 0.05$. For each DM mass, we
find the values of $A_\pm$, $P_\pm$ and $\tau$ which best fit the
data. Then we constrain the DM decay lifetime at the 2$\sigma$ level
with respect to the best fit point. We show a benchmark $\mathcal
{O}_{QLD}$ model point which is constrained at the 2$\sigma$ level
in Fig.~\ref{fig:comparepppm}.

\section{Constraints on ADM decay}
\label{sec:ADMresult}
\subsection{Light ADM Scenario} \label{sec:LADM}

We begin with constraints on ADM particles with mass in the natural
window, around 10 GeV.
We take the flavor universal scenario for both $\mathcal {O}_{LLE}$
and $\mathcal {O}_{QLD}$ operators, while for $\mathcal {O}_{UDD}$,
we take both the heavy and light flavor structure, $\mathcal
{O}_{UDD_L}$ and $\mathcal {O}_{UDD_H}$, as discussed in
Table~\ref{structures}. This choice aims to illustrate the effects
of final state quark kinematics including the $b$-quark threshold
effect.

\begin{figure}
\begin{center}
\hspace*{0.15cm}
\includegraphics[width=0.75\textwidth]{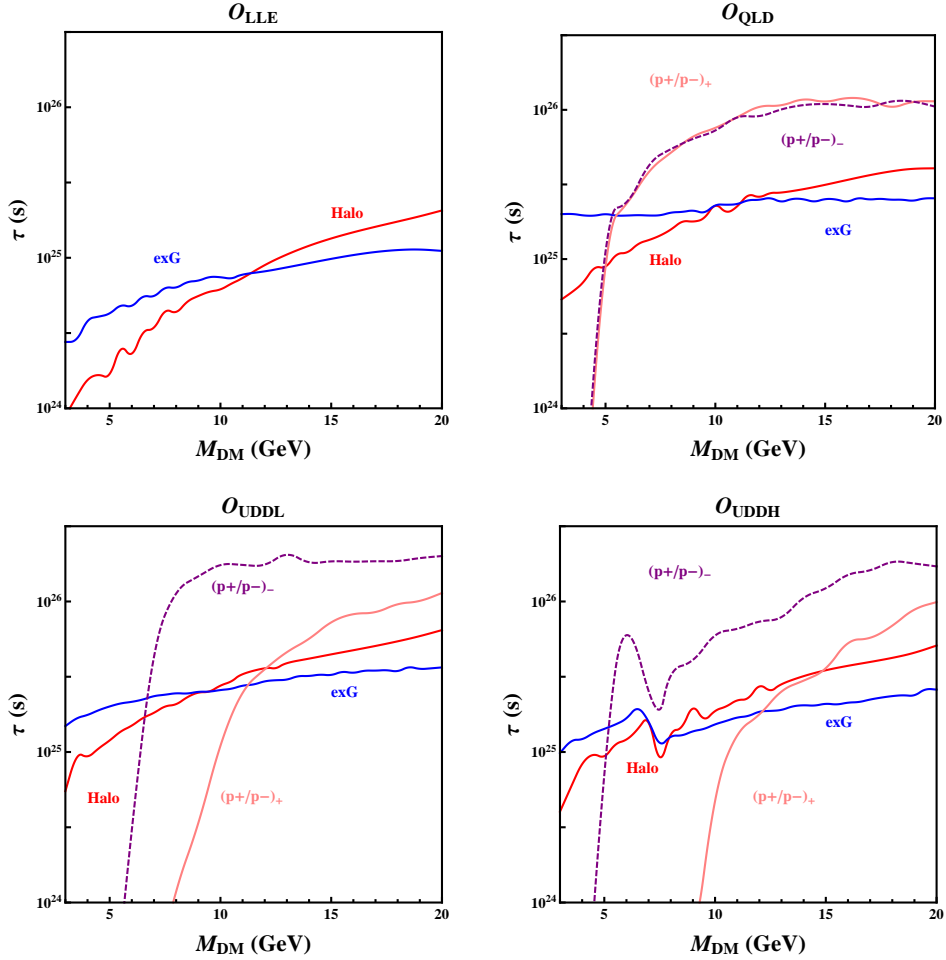}
\caption{Constraints on the lifetime $\tau$ of the DM from gamma ray spectra and from
proton/anti-proton fluxes.  We consider both scenarios where the
ADM particle carries positive or negative baryon/lepton number. As expected, the
sign of baryon number is important for the constraints
from $p+/p-$.}
\label{fig:FermiLow}
\end{center}
\end{figure}

As discussed in previous sections, we derive constraints on
light ADM decay by gamma ray spectrum and proton/anti-proton fluxes.
In Fig. \ref{fig:FermiLow}, we present our results. For each
operator, we overlay the constraints from gamma ray spectra with
those from proton/anti-proton fluxes.

For the constraints from gamma ray spectra, the constraints
are stronger when there are more hadronic particles in the final
state, as expected. The constraints on $\mathcal {O}_{UDD_L}$ are
universally stronger than the constraints on $\mathcal {O}_{UDD_H}$,
since quarks from $\mathcal {O}_{UDD_L}$ have larger kinetic
energy.

For proton/anti-proton fluxes, the constraints are very
different when DM carries positive or negative baryon number.  When
DM carries negative baryon number, there is at least one anti-proton
in the decay final states.  As illustrated in Fig.
\ref{fig:compareepem}, the anti-proton fraction is about
$10^{-5}\sim 10^{-4}$.  This is sensitive to the number of
anti-protons injected by DM decay, which gives a much stronger
constraint on decay lifetime when DM carries negative baryon number.
On the other hand, the lepton number carried by DM particles does
not make a difference for $p+/p-$ fluxes.  Thus the constraints for
$\mathcal {O}_{QLD}$ from $p+/p-$ are the same.

There are discontinuities in the constraints of $\mathcal
{O}_{UDD_H}$, both for gamma ray spectra and $p+/p-$ fluxes. The
discontinuities show up at around 7 GeV. This is caused by the
change of final state kinematics due to the open of b-quark decay
channel.

\subsection{Heavy ADM scenario} \label{sec:HADM}

The goal of this section is to show both how gamma- and charged
cosmic rays constrain heavy ADM (with mass between 100 GeV and 10
TeV), and how heavy ADM decay may generate the rising feature of the
positron-to-electron ratio observed in PAMELA and AMS-02.

\begin{figure}
\begin{center}
\hspace*{-0.75cm}
\includegraphics[width=0.80\textwidth]{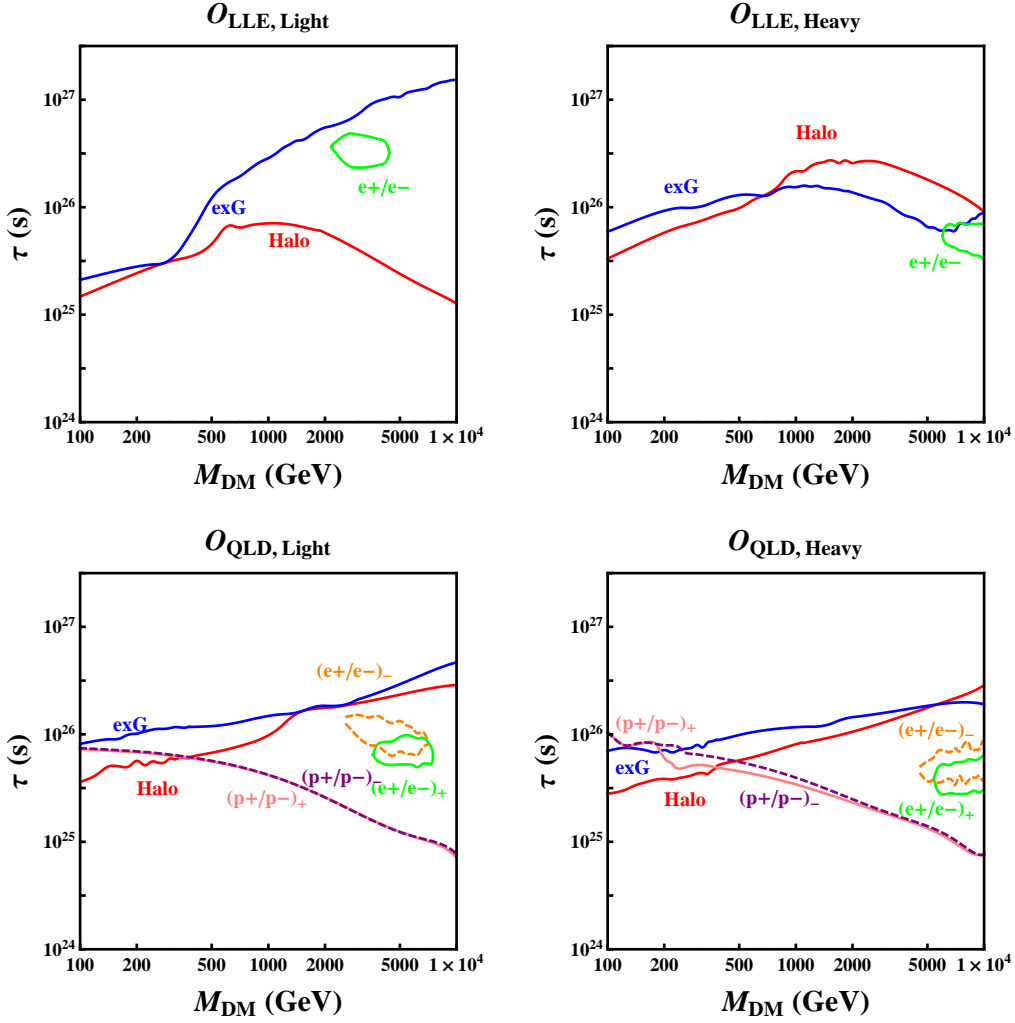} \hspace{0.1cm}
\caption{Combination of constraints and best fit regions for
$\mathcal{O}_{LLE}$ and $\mathcal{O}_{QLD}$ operators.  As discussed
in the text, constraints ({\em i.e.} the lower limits on DM decay
lifetime) are placed at 2-$\sigma$ level, while the best fit
ellipses for AMS-02 and H.E.S.S. electron/positron data
are shown at 3-$\sigma$. The left panels assume the decay products
prefer the lightest generation, while the right panels assume the
heaviest generation kinematically available is favored. We overlay
the results for ADM with positive/negative B(L) number for charged
cosmic ray fluxes, while the gamma ray constraints are the same for
these two cases. The solid lines of charged cosmic rays are for
positive B(L) number and the dashed lines are for negative B(L)
number.} \label{fig:summary1}
\end{center}
\end{figure}

\begin{figure}
\begin{center}
\hspace*{-0.75cm}
\includegraphics[width=0.80\textwidth]{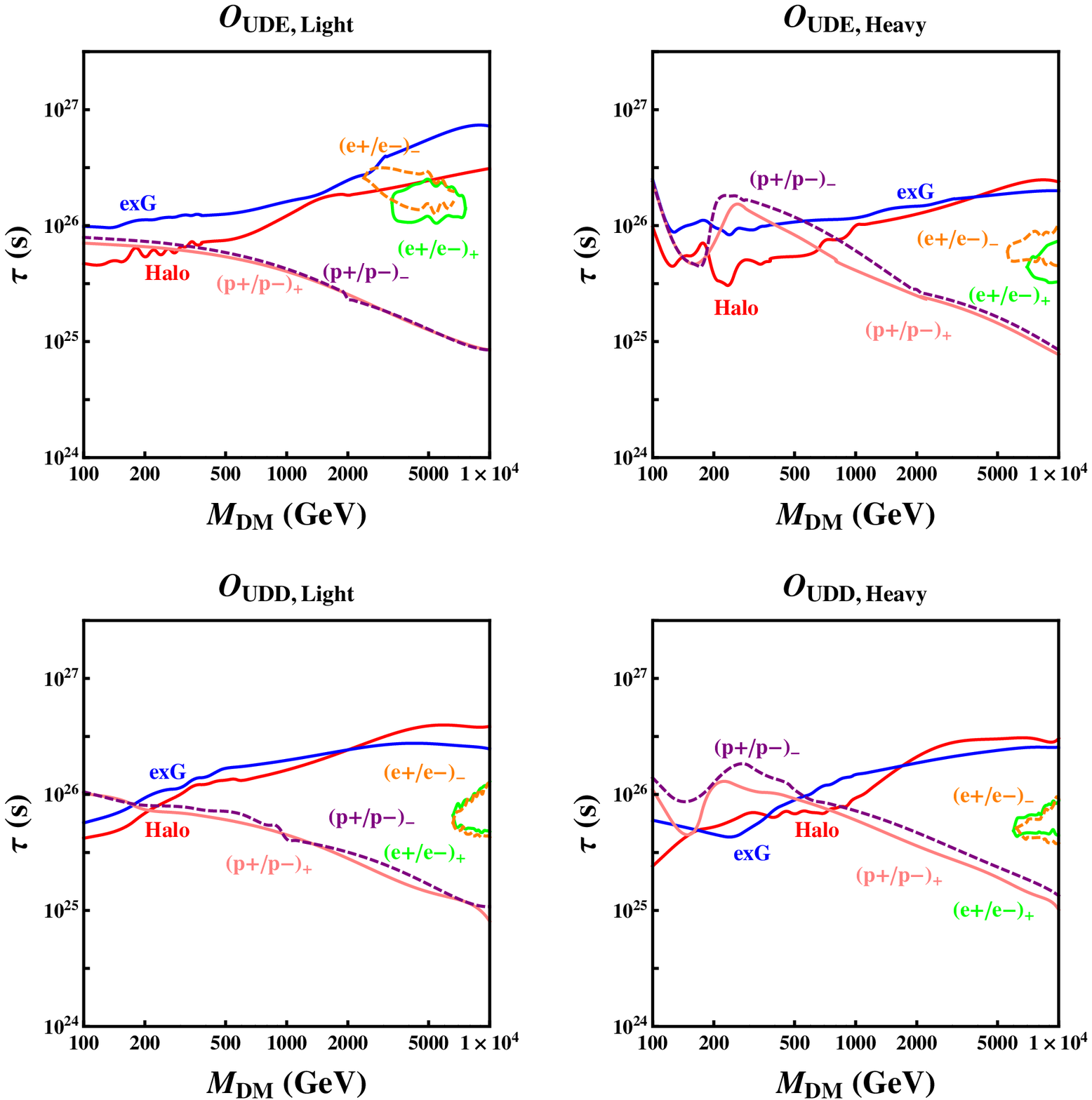} \hspace{0.1cm}
\caption{Combination of constraints and best fit regions for
$\mathcal{O}_{UDE}$ and $\mathcal{O}_{UDD}$ operators.  As discussed
in the text, constraints ({\em i.e.} the lower limits on DM decay
lifetime) are placed at 2-$\sigma$ level, while the best fit
ellipses for AMS-02 and H.E.S.S. electron/positron data
are shown at 3-$\sigma$. The left panels assume the decay products
prefer the lightest generation, while the right panels assume the
heaviest generation kinematically available is favored. We overlay
the results for ADM with positive/negative B(L) number for charged
cosmic ray fluxes, while the gamma ray constraints are the same for
these two cases. The solid lines of charged cosmic rays are for
positive B(L) number and the dashed lines are for negative B(L)
number.} \label{fig:summary2}
\end{center}
\end{figure}

\begin{figure}
\begin{center}
\hspace*{-0.75cm}
\includegraphics[width=0.40\textwidth]{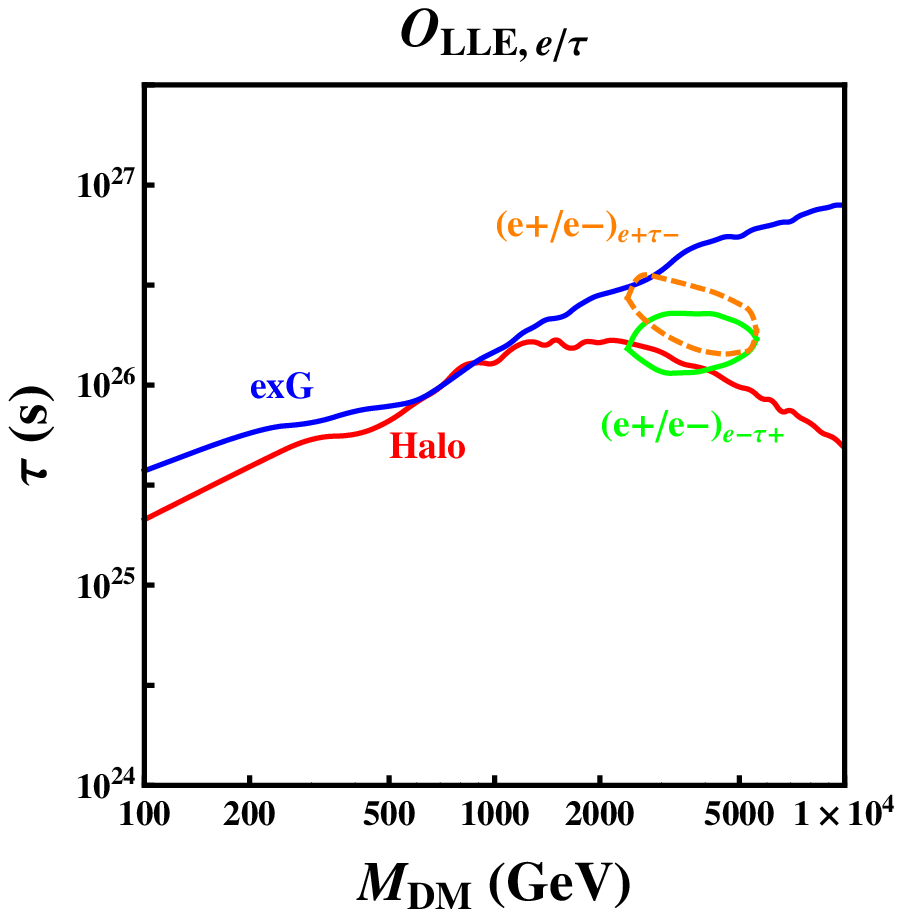} \hspace{0.1cm}
\caption{Combination of constraints and best fit regions for
$\mathcal{O}_{LLE}$ operator with the flavor asymmetric decay $X
\rightarrow e^\pm +\tau^\mp +\nu (\bar{\nu})$. As discussed in the
text, constraints are placed at 2-$\sigma$ level, while the best fit
of AMS-02 and H.E.S.S. electron/positron data is shown at
3-$\sigma$. We overlay the results for ADM with positive/negative
B(L) number for charged cosmic ray fluxes, while the gamma ray
constraints are the same for these two cases. The solid line fits to
AMS-02/H.E.S.S. data are for $e^-/\tau^+$ decay while the dashed
lines are for $e^+/\tau^-$ decay.} \label{fig:ETComb}
\end{center}
\end{figure}

In the previous sections, we addressed each indirect detection
channel carefully.  Now we combine all channels for each operator to
examine in detail whether there are regimes in parameter space which
can fit AMS-02 while remaining consistent with other constraints.
For the two extremal flavor choices, {\em i.e.} lightest and
heaviest generation fermions in the final states, the combined
results are shown in Figs.~\ref{fig:summary1},~\ref{fig:summary2}.
For the flavor asymmetric decay of $\mathcal {O}_{LLE}$, i.e.
$DM\rightarrow e^\pm+\tau^\mp+\nu (\bar{\nu})$, the combination of
various channels is presented in Fig. \ref{fig:ETComb}.

As we discussed previously, ADM can carry either positive or
negative B/L number.  Obviously this does not affect the gamma ray
spectrum, but it is crucial for studies of charged cosmic rays as
can clearly be seen in Figs.~\ref{fig:summary1}-\ref{fig:ETComb}.  The
difference is obvious for $\mathcal {O}_{QLD}$ and $\mathcal
{O}_{UDE}$ operators, as well as for the flavor asymmetric decay of
$\mathcal {O}_{LLE}$, though when the decay products involve
the lightest generation, the difference is maximally enhanced on account of the hard lepton in the final state.  If $\mathcal
{O}_{QLD}$ or $\mathcal {O}_{UDE}$ carries negative lepton number,
then the hard lepton is a positron, and similarly for the asymmetric
decay of $\mathcal {O}_{LLE}$.  Since the rising feature in the positron
fraction is most sensitive to the hard positron in the final states,
this substantially affects the fits.

From Figs. \ref{fig:summary1}-\ref{fig:ETComb}, we see the
best fit regions are confined to be small ellipses.  The positron data prefers fairly heavy DM, with mass above several hundred GeV, and since the current data
for the positron ratio stops around 300 GeV, it does not impose an
upper limit on the DM mass.  On the other hand, the electron/positron
total fluxes provide further constraints on both the very low and very
high DM mass region. In particular, data from the H.E.S.S. measurement does
not connect to the AMS-02 data smoothly.  A bump appears around
1 TeV when we combine these two data sets, which imposes a
preference for a DM mass around a few TeV, as illustrated in
Fig.~\ref{fig:compareepem}.

In the ADM decay scenario, one may have hard positrons in the final
states without generating an equal number of electrons.  This helps
to reduce the energetic byproducts from the decay, including the
gamma ray flux associated with the charged leptons. Unfortunately,
when we combine our results from the electron/positron ratio with
other constraints, the preferred region is still in tension with
other measurements, especially the diffuse extra-galactic gamma ray
flux.  This is largely because the H.E.S.S. feature around 1 TeV
imposes a lower bound on the preferred DM mass.  This feature,
however, appears at the connection region between the two data sets,
which is worrisome (recall that a similar type of feature appeared
in the ATIC data at lower energy before both Fermi and AMS-02
concluded that no such feature was present). Having a better
statistics measurement of electron/positron total flux at higher
energy is thus necessary for drawing any definite conclusions from
this analysis.


As noted above in Table~\ref{structures}, and in Figs.~\ref{fig:summary1}-\ref{fig:summary2}, we chose to present two
extremal limits, decay to the heaviest or lightest generation.
Taking $\mathcal {O}_{LLE}$ as an example, comparing the two
extremal flavor cases, when the heaviest generation, {\em i.e.}
$\tau$, is preferred, the constraint from the galactic halo gamma
ray flux is much stronger on account of the photons from hadronic
$\tau$ decay. On the other hand, the electron/positron spectra are
much harder when the first generation leptons are preferred in the
decay products. This has two consequences.  First, a harder IC
contribution to the diffuse extra-galactic gamma ray flux is
present, which leads to a much stronger constraint when the DM mass
is large. Second, since the rising feature of the positron ratio in the AMS-02
data is easier to fit, a longer decay lifetime is preferred.
Although the best-fit region from the $e^\pm$ measurement
is in tension with gamma ray measurement in both scenarios, the
tension is much weaker when DM only decays to first generation
fermions. Similar arguments can also be applied to other operators
as one can see from Figs. \ref{fig:summary1}-\ref{fig:ETComb}.

When the up-type quark is involved in the final state, whether the
top quark is kinematically allowed is the most important
feature.\footnote{A similar phenomenon also appears for the light
ADM scenario when the bottom quark is involved.}  For example, for
DM decaying through the $\mathcal {O}_{UDE}$ or $\mathcal {O}_{UDD}$
operator, as can be seen in Fig. \ref{fig:summary2}, when the
heaviest generation fermions are preferred, the constraint from
proton/anti-proton fluxes around 200 GeV is not smoothly connected
to that in higher mass region. This feature around the top quark
threshold  is not as pronounced for $\mathcal {O}_{QLD}$ in Fig.
\ref{fig:summary1}, which is mainly because $\mathcal {O}_{QLD}$ has
two decay channels for third generation particles in the final
states, {\em i.e.} $DM\rightarrow\tau^- + t + \bar{d}$ and
$DM\rightarrow\nu + b +\bar{b}$.

To summarize, in the ADM decay scenario considered in this paper,
the best fit regions from the electron/positron analysis are in
tension with other measurements for all operators we consider.  This
is largely due to the rising feature in the H.E.S.S. data around 1
TeV, which needs to be further investigated with better measurements
from AMS-02 before a definite conclusion can be drawn. For $\mathcal
{O}_{LLE}$, $\mathcal {O}_{QLD}$ and $\mathcal {O}_{UDE}$, the
tension is much weaker than from $\mathcal {O}_{UDD}$, as expected.
As is well known, the flavor structure of these operators is also
crucial. If the third generation particles are dominant in the final
states, the tension is much stronger.  We also showed that whether
ADM carries positive or negative B(L) number has impact on the
signatures, providing a possible handle to probe the asymmetry
generating mechanism of ADM.

\section{Conclusions}
\label{sec:conclusion}

In this paper we have studied signatures for decaying ADM through a
higher dimension operator.  While most models of ADM in the
literature have assumed that the ADM is absolutely stable ({\em
e.g.} through a $Z_2$ symmetry or through $R$-parity), the apparent
stability of the DM may simply be due to a very high suppression
scale of the higher dimension operator.  These same higher dimension
operators, as shown in Eq.~(\ref{ADMops}), are responsible for the
asymmetry generation in the DM sector.  Thus one may be able to
connect indirect detection signatures to the ADM mechanism.  In
addition, the asymmetry in the DM sector gives unique signatures
that allow one to prove through indirect detection the {\em sign} of
the B/L number carried by the DM.

We focused on four Fermi interactions, where a suppression scale $M$
for the operator just below the GUT scale is sufficient to be
consistent with all constraints.  We considered both ADM in its
natural mass window around 10 GeV, as well as heavier ADM with mass
between 100 GeV and 10 TeV.  In the former case, we study the
constraints from both gamma ray spectra and proton/anti-proton
fluxes; generally the constrained lifetime translates to a
constraint on the suppression scale of around $10^{13} \mbox{ GeV}$.
For heavier ADM, we fit AMS-02 and H.E.S.S. data to the models and
consider constraints from high energy FERMI data as well as the
proton/anti-proton fluxes in PAMELA. In this case, a suppression
scale of around $10^{15}-10^{16}$ GeV is appropriate for fitting
AMS-02 and H.E.S.S. data.  We were able to demonstrate the effect of
the sign of the ADM B/L asymmetry on the signatures.

Determining the nature of the DM is a complex multi-faceted problem.
Further determining how the DM density is set, for example through a
cosmic asymmetry, is an even greater challenge.  Astrophysical
objects, such as stars and neutron stars can also be crucial probes,
though they give no hint as to how the asymmetry was generated in
the first place in the DM sector. (See \cite{Zurek:2013wia} and the
references therein for review.) For ADM communicating with the SM
through higher dimension operators, if the suppression scale of the
operator is between 1 TeV and $10^4$ TeV, collider and flavor
signatures are relevant for probing ADM, as explored in
\cite{Kim:2013ivd}. 
For a much higher suppression scale,
around the GUT scale, however, one may worry that determining the
nature of the ADM mechanism becomes essentially impossible. Here we
have shown that indirect detection in these cases may provide a
handle, lending one more tool in the hunt for the DM.

\subsection*{Acknowledgments}
We would like to thank Marco Cirelli, Jeremy Mardon, Michele
Papucci, Paolo Panci, Alessandro Strumia, Meng Su, Wei Xue for
helpful discussions. The work of KZ is supported by NASA
astrophysics theory grant NNX11AI17G and by NSF CAREER award PHY
1049896.  YZ is supported by ERC grant BSMOXFORD no. 228169 and NSF
grant PHY-1316699.

\appendix
\section{A Toy Model for Heavy ADM} \label{sec:HeavyADM}

In most ADM models, the DM particle's mass is naturally around a few
GeV.  Small variations in the model, however, can easily bring the
DM mass out of its natural range. In this section, we provide a toy
model for heavy ADM with the asymmetry generated via
Eq.~(\ref{ADMops}). The ADM retains its asymmetry through this
process.

Given a concrete model, once DM and SM sectors are in equilibrium,
the baryon number deposited into the DM sector is fixed.  If there is
only one component of DM, the DM mass is fixed by DM energy density
$\Omega_{DM}$. However, if there are multiple particles in the
DM sector, for example if one is heavy and one is light, the
light DM particles can carry more of the baryon number of the entire sector while the heavy DM
particles contribute dominantly to $\Omega_{DM}$.  Such a model can be
easily built, and here we present our toy model following this logic.  We assume there are two components of DM particles, $X$ and
$\phi$, and the Lagrangian for interactions in the DM sector is written
as
\begin{equation}
\mathscr{L}_{DM}= \frac{y}{2} X_L X_L \phi^* - \frac{y}{2} X^c_R
X^c_R \phi +h.c. +\frac{\lambda}{4}\phi^2
\phi^{*2},\label{Eq:toymodel}
\end{equation}
where $X_L$ and $X_R^c$ are two Weyl spinors components of $X$. $X$
carries one unit of baryon/lepton number, depending on how $X$
couples to SM sector.  $\phi$ is a complex scalar field which
carries two units of B/L number.  We assume $m_X$ is much larger
than $m_{\phi}$.


If we assume that the transfer of the SM baryon or lepton number to the DM sector decouples at a high temperature, the baryon or lepton  number in
the DM sector is locked.  
The details are highly model dependent, but the ratio of the primordial asymmetries in the two sectors is ${\cal O}(1)$.

When the temperature drops below the transfer decoupling
temperature, the interaction within the DM sector is still active.
Due to $B$ conservation, there is no 2-to-2 process (if we restrict
ourselves to marginal operators for the annihilation)\footnote{If we
instead allow the annihilation to proceed through higher dimension
operators (for example through an interaction $X X' \phi$, where
$X'$ is exchanged in the $t$-channel and is heavier than $X$),
2-to-2 annihilation $X X \rightarrow \phi \phi$ may proceed, though
suppressed by the mass scale of the particle ($X'$ here) being
integrated out.  The essential dynamics of the models we consider
below is unchanged, though some numbers will be modified.} capable
of transferring baryon number from $X$ to $\phi$. One has to rely on
a 2-to-3 process, {\em i.e.} $X+X\rightarrow 2\phi+\phi^*$.  The
scattering cross section for this process is
\begin{equation}
\sigma v \sim \frac{y^2 \lambda^2}{8192\pi^3 m_X^2},\label{Eq:2to3Xsec}
\end{equation}
which controls the abundance of $X$ in the DM sector. We
label the temperature when this 2-to-3 process freezes out as
$T_{X,\phi}$.  This is the freeze-out temperature of the chemical
equilibrium between $X$ and $\phi$.  We assume that the freeze-out
temperature for kinetic equilibrium is much lower than
$T_{X,\phi}$.  Thus both $X$ and $\phi$ are thermal,
and their number densities are described by a Boltzmann distribution
at $T_{X,\phi}$. This is a reasonable assumption, because one needs a
large annihilation cross section to deplete the symmetric
component of ADM.

If $T_{X,\phi}$ is larger than $m_X$, both $X$ and $\phi$ are
relativistic.  The asymmetries of number densities in $X$ and $\phi$
depend on the chemical potentials as
\begin{equation}
\Delta n_i =\frac{g_i
T_{DM,SM}^3}{6\pi^2}[\pi^2(\frac{\mu_i}{T_{DM,SM}})+(\frac{\mu_i}{T_{DM,SM}})^3]\simeq
\frac{g_i T_{DM,SM}^3}{6}(\frac{\mu_i}{T_{DM,SM}}).
\label{Eq:NumDenDiff}
\end{equation}
Since the chemical potentials for $X$ and $\phi$ only differ by a
factor of 2, the asymmetries carried by these two particles are
still comparable to each other. Thus the DM mass cannot be too large
to obtain the correct DM density.

If instead $T_{X,\phi} < m_X$, $X$ is non-relativistic while $\phi$
is relativistic. For non-relativistic particles, the chemical
potential is related to the number density difference as
\begin{equation}
\Delta n_i = 2 g_i (\frac{m_i T_{X,\phi}}{2\pi})^{3/2}\
\textrm{Sinh}[\mu_i/T_{X,\phi}]e^{-m_i/T_{X,\phi}}\simeq \frac{2
g_i\mu_i}{T_{X,\phi}} (\frac{m_i T_{X,\phi}}{2\pi})^{3/2}
e^{-m_i/T_{X,\phi}}.
\end{equation}
Given the fact that $m_X>T_{X,\phi}>m_\phi$, we have
\begin{equation}
a\equiv\frac{\Delta n_{X}}{\Delta n_{\phi}}|_{T_{X,\phi}} = 12
e^{-m_X/T_{X,\phi}}\frac{g_X}{g_\phi}\frac{\mu_X}{\mu_\phi}(\frac{m_X}{2\pi\
T_{X,\phi}})^{3/2}=12 e^{-m_X/T_{X,\phi}}(\frac{m_X}{2\pi\
T_{X,\phi}})^{3/2}. \label{Eq:XPhiRatio}
\end{equation}
Assuming the symmetric component of $X$ is annihilated completely
and $\phi$'s are too light to contribute significantly to the DM
energy density, then we need $a\sim 10^{-3}$ to obtain the correct
relic abundance for TeV mass of $X$. This implies
$m_X/T_{X,\phi}\sim 10$ from Eq. (\ref{Eq:XPhiRatio}).

To determine the required cross section, we compare the interaction
rate with Hubble,
\begin{equation}
\frac{n_X \sigma v}{H}|_{T_{X,\phi}}\simeq 1 \label{Eq:MPhiFO}
\end{equation}
The cross section of $X+X\rightarrow 2\phi+\phi^*$ is calculated as
Eq. (\ref{Eq:2to3Xsec}).  For $T < m_X$, $n_X=g_X (\frac{m_X
T}{2\pi})^{3/2}exp[-(m_X-\mu_X)/T]$, $H=1.66\sqrt{g_*}\ T^2/M_{pl}$.
Taking $m_X=5$ TeV as an example, to satisfy Eq. (\ref{Eq:MPhiFO}),
one needs $y^2\lambda^2\sim 10^{-4}$, which is a reasonable choice
of parameters with $y\sim \eta \sim 0.1$.

\providecommand{\href}[2]{#2}\begingroup\raggedright
\endgroup

\end{document}